\documentclass{svjour3}
\usepackage{amssymb,graphicx,amsmath,graphics,subfigure,xfrac,url}

\usepackage{epsfig}
\usepackage{dcolumn}
\usepackage{bm}
\usepackage[utf8]{inputenc}
\usepackage{graphics}
\usepackage{graphicx}
\usepackage{epsfig}
\usepackage{amssymb}
\usepackage{amsmath}
\usepackage{color}
\usepackage{bbold}
\usepackage{xfrac}

\begin{document}

\title{A relativistic position--dependent mass system of bosonic field in cosmic string space--time background}

\author{Abbad Moussa \and Houcine Aounallah \and Sebasti\'an Valladares \and Clara Rojas}

\institute {Laboratory of Applied and Theoretical Physics.
Echahid Cheikh Larbi Tebessi University, Tebessa, Algeria. \email{moussa.abbad@univ-tebessa.dz}\\ \and Department of Science and Technology.
Echahid Cheikh Larbi Tebessi University, Tebessa, Algeria. \email{houcine.aounallah@univ-tebessa.dz} \\ \and Faculdad de F\'isica, Universidad de Sevilla, 41012 -- Sevilla, Spain. \email{sebvalsan@alum.us.es} \\ \and Yachay Tech University, School of Physical Sciences and Nanotechnology, Hda. San Jos\'e s/n y Proyecto Yachay, 100119, Urcuqu\'i, Ecuador. \email{crojas@yachaytech.edu.ec}
}

\maketitle

\begin{abstract}
In this work, we investigate the relativistic quantum motions of spin--zero scalar bosons via the Duffin--Kemmer--Petiau (DKP) equation with a position--dependent mass (PDM) system in the background of the topological defect space--time produced by a cosmic string. We determine the radial wave equation and obtain the exact analytical solutions of the wave equation for the linear and Cornell--type potential through the Bi--Confluent Heun  differential equation. In fact, we have obtained the ground state energy for both potentials.
\end{abstract}



\section{Introduction}

The investigation of quantum dynamics of particles (spin$-0$, spin$-1/2$, spin$-1$ scalar and vector bosons) in various curved space backgrounds has been of growing research interest in current times \cite{guvendi:2021,bounali:2020,aounallah:2019,boumali:2018}. Many authors introduced an electromagnetic vector potential by the non--minimal substitution of the momentum vector $p_{\mu} \to (p_{\mu}-e\, A_{\mu})$ and scalar potential $S(t,r)$ by modifying the mass term via $M \to [M+S(t,r)]$ in the relativistic either Klein--Gordon wave equation or DKP equation. In addition, many authors studied position--dependent mass quantum systems defined this way also, such as $M \to 
M(r)=M_0\,f(r)$, where $f(r)$ is an arbitrary function.

The study of position--dependent mass (PDM) has become more attractive in the literature \cite{merad2007dkp}, especially due to its applications in several areas of physics, for instance, in the study of quantum dots or in the electronic properties of semiconductors. A PDM system can be created by including a potential dependence on the mass term. A common assumption is to use a scalar potential, as mentioned above. Nevertheless, it is not the only possibility.

The problem of PDM has been analyzed in both the relativistic and non--relativistic quantum systems, with several different approaches using the Klein--Gordon equation and the Dirac equation \cite{hammoud2017bound,merad2007dkp}, for both spin$-0$ and spin$-1/2$ particles. Yet, there are still unexplored areas, and the problem can be used to study different phenomena. 

The relativistic Duffin--Kemmer--Petiau (DKP) equation allows us to study the systems with the most common integer spin, specifically those with spin$-0$ and spin$-1$, with a richer background to understand the interactions mainly of the last one. This first--order relativistic equation is considered an extension of the famous Dirac equation, in which beta matrices replace the gamma matrices. These new matrices follow another commutation rule\cite{valladares2023superradiance}, which gives rise to the DKP algebra \cite{bahar2013relativistic}. 

In this context, the DKP equation can help model the behavior and interactions of spin$-0$ particles within the spacetime influenced by cosmic strings 
\cite{hosseinpour2015dkp,yang2021dkp,castro:2015}. Cosmic strings are considered topological defects that exist in the fabric of spacetime \cite{sakellariadou2009cosmic,vachaspati2015cosmic}. They have remained a fascinating subject of study in theoretical physics for many years. In the early universe model, they were initially seen as remnants of phase transitions shortly after the Big Bang \cite{hindmarsh1995cosmic}. These cosmic--scale structures could affect various astrophysical phenomena, including gravitational lensing, gravitational waves, and the cosmic microwave background \cite{vachaspati2015cosmic}. Also, in the context of G\"odel--type space--time, the DKP equation has been studied \cite{ahmed:2020}.

The purpose of this paper is to study the relativist quantum motions of spin$-0$ scalar bosons using the DKP equation with a PDM system in the background of the topological defect space--time produced by a cosmic string. Section \ref{DKP} provides the mathematical framework of the DKP equation with a PDM for spin$-0$ in cosmic string spacetime. Section \ref{subsection_linear} introduces a linear potential, obtaining the recurrence relation and energy for this potential. Section \ref{subsection_Cornell} uses a Cornell--type potential consisting of a scalar plus a Coulomb term. from which the recurrence relation and energy are also obtained. Finally, Section \ref{conclusions} has the conclusions of our work. 

\section{DKP spin$-0$ in cosmic string spacetime} 
\label{DKP}

The relativistic quantum dynamics of spin$-0$ scalar bosons of mass $m$ in curved space is described by the DKP equation \cite{dkp1,dkp2,dkp3,dkp4} given by

\begin{eqnarray}
\left[i\widetilde{\beta}^{\mu}\left(\partial_{\mu}+\frac{1}{2}\omega_{\mu ab}S^{ab}\right)-m\right] \Psi=0,
\end{eqnarray}
where $S^{ab}=\left[\beta^{a},\beta^{b}\right]$ and $\widetilde{\beta}^{\mu}=e_{(a)}^{\mu}\beta^{a}$ with ${\beta}^{\mu}$ being the DKP matrices which satisfy the following commutation rules \cite{castro:2015,boutabia:2016}

\begin{eqnarray}
\beta^{\kappa}\beta^{\nu}\beta^{\lambda}+\beta^{\lambda}\beta^{\nu}\beta^{\kappa}=g^{\kappa\nu}\beta^{\lambda}+g^{\nu\lambda}\beta^{\kappa},
\end{eqnarray}
 $g^{\mu\nu}=\text{diag}\left(1,-1,-1,-1\right)$ is the Minkowski metric tensor. 
 
The beta matrices are chosen as follows 
\cite{td3}

 \begin{eqnarray}
\beta^{0}=\left(\begin{array}{cc}
\nu & \tilde{0}\\
\tilde{0}_{T} & \mathbf{0}
\end{array}\right),\,\beta^{i}=\left(\begin{array}{cc}
\hat{0} & \rho^{i}\\
-\rho_{T}^{i} & \mathbf{0}
\end{array}\right),
\end{eqnarray}
with $\hat{0}, \tilde{0}$, $\mathbf{0}$ as $2\times2$, $2\times3$, $3\times3$ zero matrices, respectively. Letter $T$ means transposed of $\tilde{0}$ matrix, and  $\rho$ matrix, being

\begin{eqnarray}
\nu=\left(\begin{array}{cc}
0 & 1\\
1 & 0
\end{array}\right),\,\rho^{1}=\left(\begin{array}{ccc}
-1 & 0 & 0\\
0 & 0 & 0
\end{array}\right),\,\rho^{2}=\left(\begin{array}{ccc}
0 & -1 & 0\\
0 & 0 & 0
\end{array}\right),\,\rho^{3}=\left(\begin{array}{ccc}
0 & 0 & -1\\
0 & 0 & 0
\end{array}\right).
\end{eqnarray}

The tetrad relations and the spin connection are calculated by using the relation

\begin{eqnarray}
\omega_{\mu ab}=e_{'(a)l}e_{(b)}^{j}\Gamma_{j\mu}^{l}-e_{(b)}^{j}\partial_{\mu}e_{'(a)j},
\end{eqnarray}
where $\Gamma_{\nu\lambda}^{\mu}$ are the Christoffel symbols \cite{td5} given by

\begin{eqnarray}\label{kg3}
\Gamma_{\nu\lambda}^{\mu}=\frac{g^{\mu\rho}}{2}\left(g_{\rho\nu,\lambda}+g_{\rho\lambda,\nu}-g_{\nu\lambda,\rho}\right).
\end{eqnarray}

The cosmic string metric is 

\begin{eqnarray}
\mathrm{d}s^{2}=\mathrm{d}t^{2}-\mathrm{d}r^{2}-r^{2}\mathrm{d}\theta^{2}-a'^{2}r^{2}\sin^{2}\theta \mathrm{d}\varphi^{2},
\end{eqnarray}
where $-\infty<t<+\infty$, $0\leqslant r$ , $0\leqslant\theta\leqslant\pi$, and $0\leqslant\varphi\leqslant2\pi$, $a'=1-4\eta$ , and $\eta$ is the linear mass density of the string which it is defined in the range (0,1]. Here the tetrad $e_{(a)}^{\mu}$ is chosen to be

\begin{eqnarray}
e_{(a)}^{\mu}=\left(\begin{array}{cccc}
1 & 0 & 0 & 0\\
0 & 1 & 0 & 0\\
0 & 0 & \dfrac{1}{r} & 0\\
0 & 0 & 0 & \dfrac{1}{a'r\sin\theta}
\end{array}\right).
\end{eqnarray}

The spin connections are;

\begin{eqnarray}
\omega_{\theta ab}=\left(\begin{array}{cccc}
0 & 0 & 0 & 0\\
0 & 0 & 1 & 0\\
0 & -1 & 0 & 0\\
0 & 0 & 0 & 0
\end{array}\right),\,\omega_{\varphi ab}=\left(\begin{array}{cccc}
0 & 0 & 0 & 0\\
0 & 0 & 0 & a'\sin\theta\\
0 & 0 & 0 & a'\cos\theta\\
-a'\sin\theta & -a'\cos\theta & 0 & 0
\end{array}\right),
\end{eqnarray}
after calculations, we have that the differential equation for the  radial and the angular part is given by

\begin{eqnarray}
\label{eq_chi}
\left[ \dfrac{\mathrm{d}^{2}}{\mathrm{d}r^{2}}+\frac{2}{r}\frac{\mathrm{d}}{\mathrm{d}r}+E^{2}-\dfrac{l\left(l+1\right)}{r^{2}}\right]\chi\left(r\right)=\left[m+S\left(r\right)\right]^{2}\chi\left(r\right),
\end{eqnarray}
with $l = 0, \pm 1, \pm 2, \pm 3, \dots$, and

\begin{eqnarray}
L^{2}=\left[\dfrac{1}{\sin\theta}\dfrac{\mathrm{d}}{\mathrm{d}\theta}\left(\sin\theta\frac{\mathrm{d}}{\mathrm{d}\theta}\right)+\dfrac{1}{a'^{2}\sin^{2}\theta}\frac{\mathrm{d}^{2}}{\mathrm{d}\varphi^{2}}\right].
\end{eqnarray}

\section{Linear Potential}
\label{subsection_linear}

\subsection{Wave Function}

For the linear potential, the function $S(r)$ has the following dependence:

\begin{eqnarray}
\label{linear}
S\left(r\right)=Cr.
\end{eqnarray}

Using the potential given by Eq. \eqref{linear}, we obtain that Eq. \eqref{eq_chi} takes the form

\begin{eqnarray}
\label{eq_linear}
\dfrac{\mathrm{d}^{2}\chi\left(r\right)}{\mathrm{d}r^{2}}+\dfrac{2}{r}\frac{\mathrm{d} \chi\left(r\right)}{\mathrm{d}r}+E^{2}\chi\left(r\right)-\dfrac{l\left(l+1\right)}{r^{2}}\chi\left(r\right)-\left(m+C r\right)^{2}\chi\left(r\right)=0,
\end{eqnarray}

In order to solve Eq. \eqref{eq_linear} we proposed the following solution

\begin{eqnarray}
\chi\left(r\right)=r^{\frac{\sqrt{1+4l(l+1)}-1}{2}}e^{-\frac{Cr^{2}+2mr}{2}}R\left(r\right),
\end{eqnarray}
making the change or variable $x=\sqrt{C}r$ we obtain the  Bi--Confluent Heun differential equation \cite{boumali:2018,vieira:2015} for the function $R(r)$

\begin{eqnarray}
\label{eq_linear_HeunB}
\nonumber
\dfrac{\mathrm{d}^{2}R\left(x\right)}{\mathrm{d}x^{2}}-\dfrac{\left(\beta_1 x+2 x^2-\alpha_1-1\right)}{x}\dfrac{\mathrm{d} R\left(x\right)}{\mathrm{d}x}-\dfrac{\left[\left(2\alpha_1-2\gamma_1+4\right)x+\beta_1\alpha_1+\beta_1+\delta_1\right]}{2x}R\left(x\right)=0,\\
\end{eqnarray}
where $R(x)$ is the Bi--Confluent Heun  function, and

\begin{eqnarray}
\alpha_1 &=& \sqrt{1+4l\left(l+1\right)},\\
\beta_1 &=& \dfrac{2m}{\sqrt{C}},\\
\gamma_1 &=& \dfrac{E^2}{C},\\
\delta_1 &=& 0.
\end{eqnarray}

The solution of Eq. \eqref{eq_linear_HeunB} is given by \cite{vieira:2015,boumali:2018}

\begin{equation}
R\left(x\right)=\textnormal{HeunB}\left(\alpha_1,\beta_!,\gamma_1,\delta_1;x\right).
\end{equation}

\subsection{Energy}

We consider the Bi--Confluent Heun function in the following power series form \cite{aounallah:2019,aounallah:2022}

\begin{eqnarray}
\label{series_linear}
R\left(x\right)=\sum_{j=0}^{\infty}a_{j}x^{j}.
\end{eqnarray}

Doing the substitution of Eq. \eqref{series_linear} into the Bi--Confluent Heun  differential equation, Eq. \eqref{eq_linear_HeunB},  we obtain the recurrence relation

\begin{eqnarray}
\label{recurrence}
a_{j+2}=\left\{\dfrac{m\left(1+\sqrt{1+4l(l+1)}\right)+2m\left(j+1\right)}{\sqrt{C}\left[j+2\right)\left[j+2+\sqrt{1+4l(l+1)}\right]}\right\}a_{j+1}\nonumber 
\end{eqnarray}
\begin{eqnarray}
-\left\{\dfrac{\frac{E^{2}}{C}-\sqrt{1+4l(l+1)}-2-2j}{\left(j+2\right)\left[j+2+\sqrt{1+4l(l+1)}\right]}\right\}a_{j},
\end{eqnarray}
with the coefficient

\begin{eqnarray}
\label{linear_a1a0m}
a_{1}=\frac{m}{\sqrt{C}}a_{0},
\end{eqnarray}

If $j=0$ in equation \eqref{recurrence},  we have 

\begin{eqnarray}
\label{linear_a2}
a_{2}=\left\{\frac{m\left(1+\sqrt{1+4l(l+1)}\right)+2m}{2\sqrt{C}\left[2+\sqrt{1+4l(l+1)}\right]}\right\}a_{1}-\left\{\frac{\frac{E^{2}}{C}-\sqrt{1+4l(l+1)}-2}{2\left[2+\sqrt{1+4l(l+1)}\right]}\right\}a_{0},
\end{eqnarray}
for a polynomial of first degree $(n = 1)$, we have that

\begin{eqnarray}
\label{an+1}
a_{n+1}=a_{2}=0.
\end{eqnarray}

For the equation \eqref{linear_a2}   we obtain the dependence of the energy $E$ with $l$ and $n$

\begin{eqnarray}
\frac{E^{2}}{C}-\sqrt{1+4l(l+1)}-2=2n,
\end{eqnarray}

and finally

\begin{eqnarray}
\label{linear_E1}
E=\pm\sqrt{C\left[2n+\sqrt{1+4l(l+1)}+2\right]}.
\end{eqnarray}

For the equation \eqref{linear_a2} 

\begin{eqnarray}
\label{linear_a1a0}
\left\{\frac{m\left[1+\sqrt{1+4l(l+1)}\right]+2m}{2\sqrt{C}\left[2+\sqrt{1+4l(l+1)}\right]}\right\}a_{1}-\left\{\frac{\frac{E^{2}}{C}-\sqrt{1+4l(l+1)}-2}{2\left[2+\sqrt{1+4l(l+1)}\right]}\right\}a_{0}=0.
\end{eqnarray}

Solving Eq. \eqref{linear_a1a0}, and using relation Eq. \eqref{linear_a1a0m} we have that

\begin{eqnarray}
\label{E2}
E^{2}=m^{2}\left[3+\sqrt{1+4l(l+1)}\right]+C\left[2+\sqrt{1+4l(l+1)}\right],
\end{eqnarray}
which by comparison with Eq.  \eqref{linear_E1} we have 

\begin{eqnarray}
C_{1,l}=\dfrac{m^{2}}{2}\left[3+\sqrt{1+4l(l+1)}\right].
\end{eqnarray}

Finally, we write the energy $E_{1,l}$ as the form

\begin{eqnarray}
E_{1,l}=\pm m\sqrt{\dfrac{1}{2}\left[3+\sqrt{1+4l(l+1)}\right]\left[4+\sqrt{1+4l(l+1)}\right]}.
\end{eqnarray}

\section{Cornell-type Potential}
\label{subsection_Cornell}

\subsection{Wave Function}

For the Cornell-type Potential the function $S(r)$ has the dependence:

\begin{eqnarray}
S\left(r\right)=C r+\dfrac{\lambda}{r}.
\end{eqnarray}

For this potential, Eq. \eqref{eq_chi} takes the form

\begin{eqnarray}
\label{eq_Cornell}
\dfrac{\mathrm{d}^{2}\chi\left(r\right)}{\mathrm{d}r^{2}}+\dfrac{2}{r}\dfrac{\mathrm{d} \chi\left(r\right)}{\mathrm{d}r}+E^{2}\chi\left(r\right)-\dfrac{l\left(l+1\right)}{r^{2}}\chi\left(r\right)-\left(m+C r+\dfrac{\lambda}{r}\right)^{2}\chi\left(r\right)=0,
\end{eqnarray}

To solve Eq. \eqref{eq_Cornell} we proposed the following solution

\begin{eqnarray}
\chi\left(r\right)=r^{\frac{\sqrt{1+4\lambda^{2}+4l(l+1)}-1}{2}}e^{-\frac{Cr^{2}+2mr}{2}}R\left(r\right).,
\end{eqnarray}
making the change or variable $x=\sqrt{C}r$, we obtain the  Bi--Confluent Heun  differential equation \cite{boumali:2018,vieira:2015}

\begin{eqnarray}
\label{eq_Cornell_HeunB}
\nonumber
\dfrac{\mathrm{d}^{2}R\left(x\right)}{\mathrm{d}x^{2}}-\dfrac{\left(\beta_2 x+2 x^2-\alpha_2-1\right)}{x}\dfrac{\mathrm{d} R\left(x\right)}{\mathrm{d}x}-\dfrac{\left[\left(2\alpha_2-2\gamma_2+4\right)x+\beta_2\alpha_2+\beta_2+\delta_2\right]}{2x}R\left(x\right)=0,\\
\end{eqnarray}
where $R(x)$ is the Bi--Confluent  Heun function and,

\begin{eqnarray}
\alpha_2 &=& \sqrt{1+4\lambda^{2}+4l(l+1)},\\
\beta_2 &=& \dfrac{2m}{\sqrt{C}},\\
\gamma_2 &=& \dfrac{E^2-2 C\lambda}{C},\\
\delta_2 &=& \dfrac{4\lambda_2 m}{\sqrt{C}}.
\end{eqnarray}

The solution of Eq. \eqref{eq_Cornell_HeunB} is given by the Bi--Confluent Heun  function \cite{vieira:2015,boumali:2018}

\begin{equation}
R\left(x\right)=\textnormal{HeunB}\left(\alpha_2,\beta_2,\gamma_2,\delta_2;x\right).
\end{equation}

\subsection{Energy}

As the case of linear potential,  we consider here the Bi--Confluent Heun  function in the  power series form \cite{aounallah:2019,aounallah:2022}

\begin{eqnarray}
R\left(x\right)=\sum_{j=0}^{\infty}a_{j}x^{j},
\end{eqnarray}
and substituting the series form for $R(x)$ into the differential equation Eq. \eqref{eq_Cornell_HeunB}, we have the recurrence relation

\begin{eqnarray}
\label{Cornell_recurrence}
a_{j+2}=m\left\{\frac{2\lambda+\left[1+\sqrt{1+4\lambda^{2}+4l(l+1)}\right]+2\left(j+1\right)}{\sqrt{C}\left(j+2\right)\left(j+2+\sqrt{1+4\lambda^{2}+4l(l+1)}\right]}\right\}a_{j+1}\nonumber 
\end{eqnarray}
\begin{eqnarray}
-\left\{\frac{\frac{E^{2}-2C\lambda}{C}-\sqrt{1+4\lambda^{2}+4l(l+1)}-2-2j}{\left(j+2\right)\left[j+2+\sqrt{1+4\lambda^{2}+4l(l+1)}\right]}\right\}a_{j},
\end{eqnarray}
with the coefficient

\begin{eqnarray}
\label{Cornell_a1a0m}
a_{1}=m\left\{\dfrac{2\lambda+\left[1+\sqrt{1+4\lambda^{2}+4l(l+1)}\right]}{\sqrt{C}\left[1+\sqrt{1+4\lambda^{2}+4l(l+1)}\right]}\right\}a_{0}.
\end{eqnarray}

In equation \eqref{Cornell_recurrence} if $j=0$ we have 

\begin{eqnarray}
\label{Cornell_a2}
\nonumber
a_{2}=m\left[\frac{2\lambda+\left(3+\sqrt{1+4\lambda^{2}+4l(l+1)}\right)}{2\sqrt{C}\left(2+\sqrt{1+4\lambda^{2}+4l(l+1)}\right)}\right]a_{1}-\left[\frac{\frac{E^{2}-2C\lambda}{C}-\sqrt{1+4\lambda^{2}+4l(l+1)}-2}{2\left(2+\sqrt{1+4\lambda^{2}+4l(l+1)}\right)}\right]a_{0},\\
\end{eqnarray}
for a polynomial of first degree $(n = 1)$, we have that

\begin{eqnarray}
\label{Cornell_an+1}
a_{n+1}=a_{2}=0.
\end{eqnarray}

For the equation \eqref{Cornell_a2} we have  

\begin{eqnarray}
\frac{E^{2}-2C\lambda}{C}-\sqrt{1+4\lambda^{2}+4l(l+1)}-2=2n,
\end{eqnarray}

and

\begin{eqnarray}
\label{Cornell_E1}
E_{n}=\pm\sqrt{C\left[2n+\sqrt{1+4\lambda^{2}+4l(l+1)}+2+2\lambda\right]}.
\end{eqnarray}

For the equation \eqref{Cornell_a2} we have 

\begin{eqnarray}
\label{Cornell_a1a0}
\nonumber
m\left\{\frac{2\lambda+\left[3+\sqrt{1+4\lambda^{2}+4l(l+1)}\right]}{2\sqrt{C}\left[2+\sqrt{1+4\lambda^{2}+4l(l+1)}\right]}\right\}a_{1}-\left\{\frac{\frac{E^{2}-2C\lambda}{C}-\sqrt{1+4\lambda^{2}+4l(l+1)}-2}{2\left[2+\sqrt{1+4\lambda^{2}+4l(l+1)}\right]}\right\}a_{0}=0.\\
\end{eqnarray}

Solving Eq. \eqref{Cornell_a1a0}, and using relation Eq. \eqref{Cornell_a1a0m} we have that

\begin{eqnarray}
\label{Cornell_E2}
\nonumber 
E^{2}&=&m^{2}\left[2\lambda+3+\sqrt{1+4\lambda^{2}+4l(l+1)}\right]\dfrac{\left[2\lambda+1+\sqrt{1+4\lambda^{2}+4l(l+1)}\right]}{\left[1+\sqrt{1+4\lambda^{2}+4l(l+1)}\right]}\\
&+&C\left[2\lambda+2+\sqrt{1+4\lambda^{2}+4l(l+1)}\right],
\end{eqnarray}
which by comparison with Eq.  \eqref{Cornell_E1} we have 

\begin{eqnarray}
C_{1,l}=\dfrac{m^{2}}{2}\dfrac{\left[2\lambda+3+\sqrt{1+4\lambda^{2}+4l(l+1)}\right]\left[2\lambda+1+\sqrt{1+4\lambda^{2}+4l(l+1)}\right]}{\left[1+\sqrt{1+4\lambda^{2}+4l(l+1)}\right]}.
\end{eqnarray}

\vspace{0.6cm}
From Fig. (\ref{fig: c_1,l}), we can observe that the coefficient $C_{1,l}$ describes a linear behavior, as expected from the form of the potential. 

\begin{figure}[ht]
\centering
\includegraphics[scale=0.9]{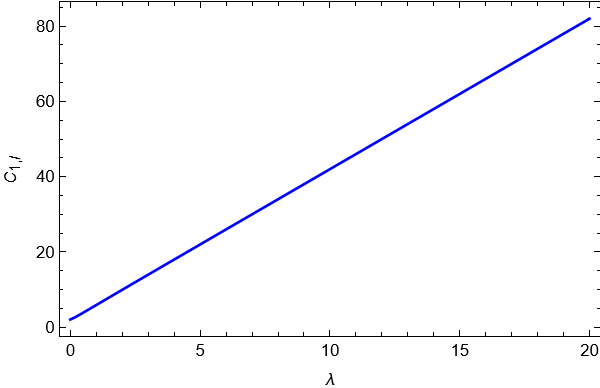}
\caption{Plot of the change of the coefficient $C_{1,l}$ as a function of $\lambda$. }
\label{fig: c_1,l}
\end{figure}

Finally, we write the energy $E_{1,l}$ as the form

\begin{eqnarray}\label{Cornell_energy}
\nonumber
E_{1,l}&=&\pm m\sqrt{\frac{\left[2\lambda+3+\sqrt{1+4\lambda^{2}+4l(l+1)}\right]\left[2\lambda+1+\sqrt{1+4\lambda^{2}+4l(l+1)}\right]}{2\left[1+\sqrt{1+4\lambda^{2}+4l(l+1)}\right]}}\\
&\times&\dfrac{\sqrt{\left[2\lambda+4+\sqrt{1+4\lambda^{2}+4l(l+1)}\right]}}{{2\left[1+\sqrt{1+4\lambda^{2}+4l(l+1)}\right]}}.
\end{eqnarray}

\begin{figure}[ht]
\centering
\includegraphics[scale=0.7]{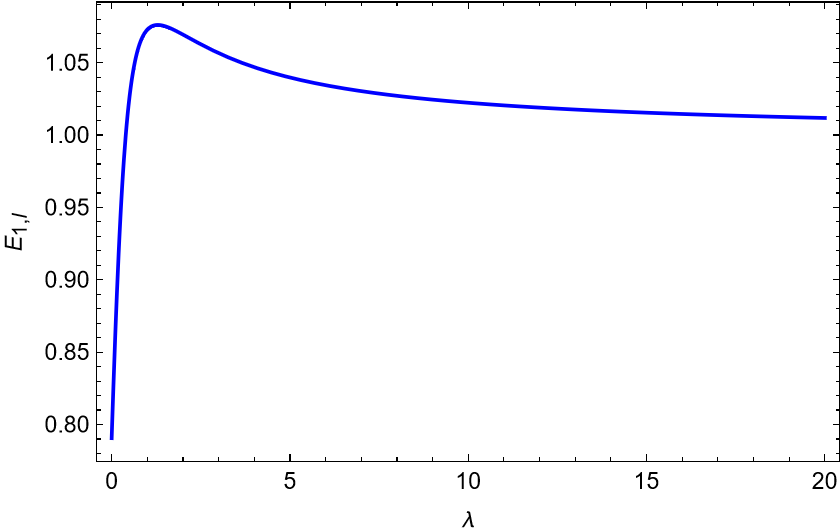}
\caption{Plot of the ground state energy level $E_{1,l}$ as a function of $\lambda$. }
\label{fig: energy2}
\end{figure}

The dependence of the energy, Eq. \eqref{Cornell_energy}, with respect to $\lambda$, can be seen in Fig. \eqref{fig: energy2}. Note that the Cornell--type potential reduces to the linear potential for $\lambda=0$.

\section{Conclusions}
\label{conclusions}

In this study, we have investigated the behavior of spin--0 scalar bosons in the presence of cosmic strings' topological defects and spacetime curvature. We used the DKP equation within a position--dependent mass framework and employed linear and Cornell--type potentials. We derived energy expressions and recursive relations by the Bi--Confluent Heun differential equation. The results show how position--dependent masses and cosmic strings' topological defects influence the system's behavior. In particular, in the case of the Cornell potential, which consists of a scalar and a Coulomb term. We observed that the Coulomb term is responsible for short-distance interactions. Moreover, in the limiting case in which $\lambda =0$ it is reduced to the scalar case.

\bibliographystyle{unsrt}

\end{document}